\documentclass[12pt]{article}
\usepackage{graphicx}
\newcommand{\be}{\begin{equation}}
\newcommand{\ee}{\end{equation}}
\newcommand{\ba}{\begin{eqnarray}}
\newcommand{\ea}{\end{eqnarray}}
\newcommand{\no}{\nonumber \\}
\begin{document}

\begin{titlepage}
\pagestyle{empty}
\vspace{1.0in}
\begin{flushright}
%\today
\end{flushright}
\vspace{.1in}
\begin{center}
\begin{large}
{\bf  Notes on D-instanton correction to $AdS_5 \times S^5$ geometry}
\end{large}
\vskip 0.5in
Chanyong Park$^{+}$ and Sang-Jin Sin$^{*}$ \\
\vskip 0.2in
{\small {\it Department of Physics, Hanyang University\\
Seoul, 133-791,  Korea}}
\end{center}

\begin{abstract}
We show that
the  D-instanton in the  $AdS_5 \times S^5$
background  is a wormhole connecting the background
$AdS_5 \times S^5$  to  the flat space
$\bf R^{10}$ located at the position of the D-instanton.
By a $SL(2,{\bf R})$ rotation of type IIB theory, we can make the {\it global}
geometry flat in string frame.
We also find that, due to the tight relation  between the dilaton and
the axion, there is no $SL(2,{\bf R})$ element that takes strong string
coupling  to weak  one  without making the axion ill defined.
We also  discuss the case of  $AdS_3$ as well as the instanton gases.
A subtlety on the D-instanton at the boundary   or at the horizon is
discussed.
\end{abstract}

\vspace{.3in}
\noindent keywords: D-brane, Instanton, String \\

\vspace{5cm}
\noindent $^{+}$ :  chanyong@dirac.hanyang.ac.kr \\
$^{*}$ :  sjs@dirac.hanyang.ac.kr
\end{titlepage}

Recently, the proposal of Maldacena \cite{malda} on the duality
between the string theory in an anti de Sitter(AdS) space and the
conformal field theory at the boundary  brought an explosive
interests, and by now refinements and significant
evidences\cite{klebanov,witten1,witten2,malda2,ReYe,imss,GO,COOT}
were established.  The correspondence open the possibility to study
non-perturbative QCD in four dimension in terms of the
semi-classical supergravity. More recently, the question of
instanton effects in these contexts
\cite{banks,ho,bianchi,ko,bergshoeff} was raised and it was shown
that the D-instanton in the $AdS_5$ background is relevant to the
instanton of the super Yang-Mills theory living at the boundary.

In this paper, we point out that the metric for the D-instanton
inside the bulk of $AdS_5 \times S^5$ background \cite{bianchi} is
a wormhole between background $AdS_5 \times S^5$ and the {\it flat
space } ${\bf R^{10}}$ located at the position of the D-instanton.
One can make the  result more surprising  by using the $ SL(2,{\bf
R})$ symmetry of the type IIB theory. In fact we will show that one
can make the   geometry  globally flat by an appropriate $SL(2,{\bf
R})$ rotation. On the other hand, the dilaton diverge at the
D-instanton position.   Ordinary recipe for this is going
 to the S-dual picture by duality rotation.
 However, we  will see that,
 due to the tight relation  between the dilaton and the axion,
 there is no $SL(2,{\bf R})$ element that rotate strong string
 coupling  to weak  one  without making the axion diverge.

\vskip.5cm

To set the notation and for the later use, we give a brief review on
D-instanton solution of type IIB supergravity in the flat background \cite
{Dinstanton} as well as that in the $AdS_5 \times S^5$ background.
We start by D-instanton in flat background.
Here we only consider the Ramond-Ramond (RR) pseudo-scalar axion $\chi $, the
dilaton $\phi$ and the metric $g_{\mu \nu }$. The bosonic part of
type IIB action in Einstein frame is given by
\begin{equation}
S_{M}=\int d^{10}x \sqrt{-g} \left[ R-\frac{1}{2}(\partial \phi )^2
      - \frac{1}{2} e^{2\phi} (\partial \chi )^2  \right],
\end{equation}
in Minkowski space. After the Wick rotation, the Euclidean action reads
\begin{equation}
S_{E}=\int d^{10}x \sqrt{-g} \left[ R -\frac{1}{2}(\partial \phi )^2
      + \frac{1}{2} e^{2\phi} (\partial \chi )^2  \right] .
\end{equation}
Notice that $\chi \rightarrow i\chi$ under the Wick rotation,
since $\chi$ is a pseudo-scalar. The equations of motion are given by
\begin{eqnarray}
0 &=& R_{\mu \nu } -\frac{1}{2} \partial_{\mu} \phi \partial_{\nu} \phi
      + \frac{1}{2} e^{2\phi} \partial_{\mu} \chi \partial_{\nu} \chi
      ,  \nonumber \\
0 &=&\partial _{\mu }(\sqrt{g}g^{\mu \nu }e^{2\phi }\partial _{\nu }\chi ),
\nonumber \\
0 &=&e^{2\phi }(\partial \chi )^{2}+\frac{1}{\sqrt{g}}\partial _{\mu }(\sqrt{%
g}g^{\mu \nu }\partial _{\nu }\phi ).  \label{eq1}
\end{eqnarray}
For the flat background, $g_{\mu \nu }=\delta _{\mu \nu }$, the solution is
given by
\begin{equation}
\pm \chi +\alpha =e^{-\phi }=\frac{1}{H}  \label{sol1}
\end{equation}
where $\alpha$ is constant and $H$ is a harmonic function i.e. $\partial ^{2}H=0$%
. Assuming the spherical symmetry, it is given by
\begin{equation}
H=h+\frac{Q}{r^{8}},  \label{harmon}
\end{equation}
where $h=e^{\phi _{\infty }}$ and $Q$ is the Noether charge defined by
\begin{equation}
Q=\pm \frac{1}{8\Omega _{9}}\int_{\partial M}e^{2\phi }\partial \chi :=N_{-1}%
{\alpha ^{\prime }}^{4}
\end{equation}
where $\Omega _{9}=2\pi ^{5/2}/24$ is the volume of the nine-sphere. This
single instanton solution is evidently singular at $r=0$ in the Einstein
frame. However, in string frame the metric become
\begin{equation}
ds^{2}=\sqrt{H}[dr^{2}+r^{2}{d\Omega _{9}}^{2}]=\sqrt{hr^{4}+\frac{Q}{r^{4}}}%
\left[ \left( \frac{dr}{r}\right) ^{2}+{d\Omega _{9}}^{2}\right] .
\end{equation}
Note that the solution is invariant under the inversion transformation
\begin{equation}
r\rightarrow \left( \frac{Q}{h}\right) ^{1/4}\frac{1}{r}.  \label{inversion}
\end{equation}
It corresponds to a wormhole connecting two asymptotically flat Euclidean
regions. The wormhole-throat has minimal diameter at $r=r_{min}$ given
by
\begin{equation}
{r_{min}}^{8}= \frac{Q}{h}  .
\end{equation}
Under the inversion (\ref{inversion}), the asymptotic flat regions at $r=0$
and $r=\infty $ are mapped onto each other.

We now turn to the
D-instanton in $AdS_5\times S^5$ background.
We first  describe the background geometry as the near
horizon geometry of D3 branes. The $Dp$-brane as a supergravity solution can
be characterized in terms of $H_p (x_{\perp})$, a harmonic function of the
coordinates perpendicular to the world volume of Dp-brane. In fact, $H_p$
depends only on the radial coordinate $r=\sqrt{{x_{p+1}}^2 + \cdots + {x_9}^2%
}$ and is given by
\begin{equation}
H_p = 1 + \frac{Q_p}{r^{7-p}},
\end{equation}
where the charge $Q_p$ is
\begin{equation}
Q_p = g_s (2\pi)^{(5-p)/2} (2\pi \alpha^{\prime})^{(7-p)/2} [2\pi^{(7-p)/2}
/\Gamma((7-p)/2)]^{-1}.
\end{equation}
In string frame, the Dp-brane metric in Euclidean version is
\begin{equation}
{ds_p}^2 = {H_p}^{-1/2} ({dx_0}^2+ \cdots + {dx_p}^2 ) + {H_p}^{1/2} ( {%
dx_{p+1}}^2 + \cdots + {dx_9}^2 )
\end{equation}
with the dilaton field $\phi$ given by
\begin{equation}
e^{2\phi} = {H_p}^{(3-p)/2}.
\end{equation}
Therefore the string background describing $N$ D3-branes is given by
\begin{equation}
{ds_3}^2 = {H_3}^{-1/2} {d\vec{x}}^2 + {H_3}^{1/2} (dr^2 + r^2 {d\Omega_5}^2
).
\end{equation}
In the decoupling limit ($\alpha^{\prime} \rightarrow 0$, $u=\frac{r}{%
\alpha^{\prime}}={\rm {fixed}}$), the constant term in the harmonic function
$H_3$ can be neglected. After rescaling the $u \rightarrow \lambda^{-1} u$
and $\vec{x} \rightarrow \lambda \vec{x}$ by a constant factor $\lambda^4 =
4\pi gN$, the metric can be written
\begin{equation}
{ds_3}^2 = \alpha^{\prime} \sqrt{4\pi gN} \left[ u^2 {d\vec{x}}^2 +\frac{du^2%
}{u^2} + {d\Omega_5}^2 \right] ,
\end{equation}
which is the metric of $AdS_5 \times S^5$. In terms of the variable $z=1/u$,
the metric is
\begin{equation}
{ds_3}^2 = \alpha^{\prime} \sqrt{4\pi gN} \left[ \frac{1}{z^2} ( {d\vec{x}}%
^2 + dz^2 + z^2{d\Omega_5}^2 )\right] .  \label{adsmetric}
\end{equation}
The boundary of $AdS_5$ is at $u=\infty$ or equivalently at $z=0$.

Now we consider D-instanton in the $AdS_5 \times S^5$ background.
The equations of motion in $AdS$ background are equal to (\ref{eq1}),
except that the first equation in (\ref{eq1}) is replaced \cite{ho} by
\be
0 = R_{\mu \nu } -\frac{1}{2} \partial_{\mu} \phi \partial_{\nu} \phi
      + \frac{1}{2} e^{2\phi} \partial_{\mu} \chi \partial_{\nu} \chi
      - \frac{1}{6} {F_{\mu}}^{\delta \rho \lambda \sigma}
         F_{\nu \delta \rho \lambda \sigma}
\ee
where $F_{\mu \nu \rho \lambda \sigma}$ is a self-dual gauge field
strength. There are two different solutions of D-instanton in the
bulk of the $AdS_5$. The first solution is pointlike in $AdS_5$ and
uniformly spread over $S^5$\cite{ho} while the second
is pointlike in ten dimensional $AdS_5 \times S^5$ space \cite{bianchi}.
These two solutions agree only at the boundary of the $AdS_5$
and equal to the solution given by \cite{ko}. Here we use the solution given
in the ref.\cite{bianchi}. The solution spread over $S^5$ can be regarded as
superposition of the former. The supergravity solution for D-instanton
embedded in the bulk of the $AdS_5 \times S^5$ is given by the metric of the
following form,
\begin{equation}
{ds}^2 = {H_{-1}}^{1/2} \left[ \frac{1}{z^2} (d\vec{x}^2 + dz^2 + z^2
         d {\Omega_5}^2) \right], \label{instmetric}
\end{equation}
where we regarded $z$ as the length of the six vector $\vec{z}$ of the
inverted transverse space of D3, i.e, $\vec{z}=\vec{x_{\perp}}/u^2$. Here,
we have taken off the constant factor ($\alpha^{\prime} \sqrt{4\pi gN}$).
The associated one and five form R-R field strengths and dilaton field are
given by
\begin{eqnarray}
F^{(1)} &=& d (H_{-1})^{-1},  \nonumber \\
F^{(5)} &=& d(1/z^4) (d{x_0} \wedge \cdots \wedge d{x_3} ),  \nonumber \\
e^{\phi}& =& H_{-1}.
\end{eqnarray}
Notice that $H_{-1}$ is a harmonic function satisfying the laplace equation
in the $AdS_5 \times S^5$
\begin{equation}
\frac{1}{\sqrt{g}} \partial_{\mu} ( g^{\mu\nu} \sqrt{g} \partial_{\nu}
H_{-1} ) = 0.
\end{equation}
Here, we give some detail of the derivation of $H_{-1}$ since it does not
appear elsewhere. Since the metric (\ref{adsmetric}) is conformaly related
to the flat metric,
\begin{equation}
ds^2 = \Gamma(z) [ {d \vec{x}}^2 + dz^2 + z^2 {d\Omega_5}^2 ]
\end{equation}
with $\Gamma(z)= 1/z^2$, we look for an laplace equation in the flat metric
by changing the variable
\begin{equation}
H_{-1} (\vec{x},\vec{z}) = G(z) H(\vec{x},\vec{z}),
\end{equation}
where $G(z)$ is a function of $z$ only. Then the laplace equation for
$H_{-1}$ becomes
\begin{equation}
0 = \frac{1}{\Gamma} \big[ G \Delta_0 H + 2 \delta^{\mu\nu} \{
\partial_{\mu} G + 2 G (\partial_{\mu} \log \Gamma) \} \partial_{\nu} H + \{
\Delta_0 G + 4\delta^{\mu\nu} (\partial_{\mu} G ) (\partial_{\nu} \log
\Gamma) \} H \big]  \label{clap}
\end{equation}
where $\Delta_0$ is a laplacian operator in ten-dimensional flat
space. If $\Gamma$and $G$ satisfies the following two equations
\begin{eqnarray}
\Delta_0 G + 4\delta^{\mu\nu} (\partial_{\mu} G ) (\partial_{\nu} \log
\Gamma) &=& 0  \nonumber \\
\partial_{\mu} G + 2 G (\partial_{\mu} \log \Gamma) &=& 0,  \label{gequ}
\end{eqnarray}
then the equation (\ref{clap}) is reduced to the flat space laplace
equation:
\begin{equation}
\Delta_0 H(\vec{x},\vec{z})=0,
\end{equation}
whose solution can be readily written by
\begin{equation}
H(\vec{x},\vec{z}) = c_1+\frac{c_2}{[(\vec{x}-\vec{x}_0 )^2 + (\vec{z}-\vec{z%
}_0 )^2 ]^4 }.
\end{equation}
where $(\vec{x}_0, \vec{z}_0)$ is the location of the D-instanton. Now for
the given $\Gamma$, the solution of (\ref{gequ}) is easily found and
given by
\begin{equation}
G(z)=c_3 z^4.
\end{equation}
Therefore the solution $H_{-1}$ can be written as
\begin{equation}
H_{-1} (\vec{x},\vec{z}) = h + \frac{ N_{-1} z_0^4 z^4}{ [(\vec{x} - \vec{x}_0
)^2 + (\vec{z}-\vec{z}_0 )^2 ]^4 }.  \label{dinst}
\end{equation}
Here $z_0^4$ factor is added for the dimensional reason. Notice that the
term like $c^{\prime}(z/z_0)^4$ could be added to above solution. However,
since it can be regarded as the large $z_0$ limit of the second term, we
deleted it. A few important remarks regarding the role of $z_0$ should be
made here.

\begin{itemize}
\item  While $\vec{x}_{0}$ could be set to zero without changing the
geometry, changing $z_{0}$ changes the geometry. This is because the
presence of the D3 branes breaks the translational symmetry in the directions
transverse to them.

\item  $z_{0}$ is the only scale that appears in the near horizon geometry
where $\alpha ^{\prime }$ is taken to be zero.
This scale is introduced by Higgsing
D3 and D(-1) by separation $z_{0}=1/u_{0}$, and it can be  interpreted as
the size of the D-instanton. In fact it is interpreted as the size of the
Yang-Mills instanton of the boundary theory\cite{ho,bianchi}.

\item  The solution is $\alpha'$ independent.
This should be so since we already took the near horizon geometry.
For this it is crucial to have a scale $z_{0}$.
We discuss the problem that rises in the absence of $z_{0}$ in the next
item.

\item By considering the near horizon geometry of the  D(-1)-D3 system
(D(-1) contained in the D3),
one can get the solution $H_{-1}= 1+ c\alpha'^4/r^4= 1+cz^4$.
The problem is that $c$ is a dimensionful constant  but we do not have any
scale other than $\alpha'$.  So, $c\sim \alpha'^{-2}$  by counting the
dimension.
Then, the constant piece is negligible in the decoupling limit
and $H_{-1}\sim z^4$ .
Then  it is easy to see that
the metric (\ref{instmetric}) becomes globally flat.
Looking from the AdS point of view,
this is not surprising since in this case
the D-instanton size is   of string length which is regarded as
 infinitely large  in the decoupling limit. Namely since $u=r/\alpha'$ fixed
 and $\alpha' \to 0$,
$l_s\sim \sqrt{\alpha'} >>>r \sim \alpha' u $.
However,  it  seems to be  puzzling  from the observer inside the D3 brane
if we regards the D-instanton as a localized object inside the D3 brane.
How can a localized object change the global geometry?
There are two possible resolution to this.
The first one is to regard the solution $H_{-1}=z^4 $ not for the point
like D-instanton but for the D-instantons spread over the world
volume of D3 just as the solution in \cite{ho} for the D-instanton
spread over the $S^5$. In fact, from  ten dimensional point of view,
the power $r^{-4}$  for the harmonic function
is acceptable only if the source is four dimensional.
 In this case $z^4$ term dominate the constant piece
and the global geometry changes from $AdS_5\times S^5$
to flat space.
The second way to resolve the problem is
to regard the $1+cz^4$ as the $z_0\to \infty$
limit of (\ref{dinst}). Since $z_0$ is the D-instanton size, it corresponds
to a very large instanton.
However, notice that the $z^4$ term in this case is negligible  since
$H_{-1}\sim h+N_{-1}(z/z_0)^4 \to h$.
Similar comments can be  applied to the D-instanton
at the boundary, which can be obtained  from $1+cz^4$
by taking the inversion ${z}\to z/(z^2+\vec{x}^2)$.
It should be considered as the $z_0\to 0$ limit of
(\ref{dinst}), therefore it is a very small instanton.
Here also  the contribution of this small instanton to the geometry change
is negligible  since $H_{-1} \sim h+z^4_0 \cdot O(1) \to h$.
\end{itemize}

\vskip.5cm

Let's discuss the instanton-correction to  $AdS_{5}\times S^{5}$ geometry.
We start the discussion by  observing that near the D-instanton position
$(\vec{x}_0, \vec{z}_0)$, the string frame metric becomes flat:
\begin{eqnarray}
ds^2 &\sim & \frac{ d\vec{x}^2 + d\vec{z}^2 } {[(\vec{x}-\vec{x}_0 )^2 + (%
\vec{z}-\vec{z}_0 )^2 ]^2} \\
&=& \frac{1}{ w^4} (d w^2 + w^2 {d \Omega_9}^2) \\
&=&d \vec{X}^2,
\end{eqnarray}
where $w=\sqrt{(\vec{x}-\vec{x}_0 )^2 + (\vec{z}-\vec{z}_0 )^2 }$ is the
ten-dimensional radius and
\begin{equation}
\vec{X}= \left(\frac{\vec{x} }{w^2}, \frac{\vec{z}} {w^2}\right)= \left (
\frac{\vec{x } }{\vec{x}^2_{\perp}/u^4 +\vec{x}^2 } , \frac{\vec{x}%
_{\perp}/u^2 }{\vec{x}^2_{\perp}/u^4+\vec{x}^2} \right).
\end{equation}
Notice that $H_{-1}$ is constant far from the D-instanton, especially near
the boundary, $z\to 0$. Therefore what we have shown is that the D-instanton
solution is a wormhole solution connecting the asymptotic $AdS_5\times S^5$
space and a flat space at the D-instanton position. See. figure 1.

\begin{figure}
\begin{center}
\includegraphics[angle=0, width=100mm]{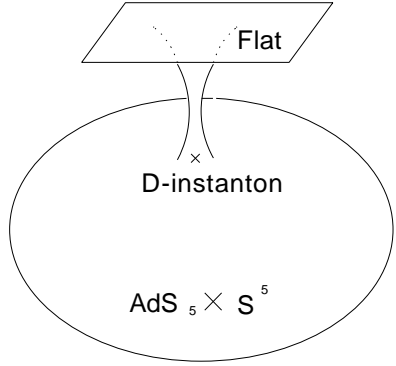}
\end{center}
\caption{ D-instanton solution is a wormhole between the AdS and
the flat space at the D-instanton position.}
\end{figure}

In fact,  one can take advantage of the $SL(2,{\bf R})$ invariance of
the Type IIB theory to make the whole geometry flat. Introducing a scalar
$S_\pm$ by
\be
S_{\pm} =\chi \pm e^{-\phi }
\ee
the action and equations of motion of type IIB theory are invariant under
the SL(2,{\bf R}) transformations
\begin{equation}
S_{\pm }\rightarrow \frac{aS_{\pm }+b}{cS_{\pm }+d}\;\;\;,\;\;\;ad-bc=1
\end{equation}
with the two generators
\begin{equation}
\Omega _{1}:S_{\pm }\rightarrow S_{\pm }+1\;\;\;,\;\;\;
\Omega _{2}: S_{\pm }\rightarrow-\frac{1}{S_{\pm }}. \label{slro}
\end{equation}
Using these transformations we can change arbitrarily the parameters $\alpha
$ and $h$ characterizing the D-instanton solution (but not $Q$). In
particular, we can transform the constant part  of the harmonic function
$H_{-1}$ to zero by any $SL(2,{\bf R})$ transformation of the form
\be
\pmatrix{ a & b \cr c & d} =
\pmatrix{ -c\alpha -\frac{h}{2c} & -c \alpha^2 +
          \frac{2c\alpha}{h} - \frac{h\alpha}{2c} \cr
          c &  c\alpha - \frac{2c}{h}},  \label{sl2r}
\ee
with arbitrary $c$ and the dilaton becomes
\begin{equation}
e^\phi = \frac{ {N_{-1}}^{\prime} z_0^4 z^4}{ [(\vec{x} - \vec{x}_0
)^2 + (\vec{z}-\vec{z}_0 )^2 ]^4 } ,  \label{gdinst}
\end{equation}
where ${N_{-1}}^{\prime}= N_{-1} (2c/h)^2$. If we require
${N_{-1}}^{\prime} = N_{-1}$, then we get $c=\pm h/2$.
Now,  the metric is flat {\it globally}. This is  surprising.
 It indicates that the special
SL(2,{\bf R}) rotation (\ref{sl2r}) effectively lead us to the near horizon
geometry of the D-instanton. One remark is that above statement is
approximation, since the SL(2,R) is broken to
SL(2,Z) due to the quantum effect.

\vskip.5cm

Next, we
observe that the instanton correction ruins the conformal invariance of the
D3 system. That is, in the super-Yang-Mills theory on the world volume of
the Dp-brane, the dimensionless effective coupling constant $g_{eff}$ is
given by \cite{imss}
\begin{equation}
{g_{eff}}^2 = N g^2_{YM}u^{p-3}.
\end{equation}
Notice that for $p=3$, $g_{eff}$ does not run as the energy scale
($u=r/\alpha^{\prime}$) changes, implying the conformal invariance. The
supergravity solution in the decoupling limit is reliable for $gN >> 1$. On
the other hand, after the D-instanton insertion, the string coupling $g_s =
e^{\phi} $ runs as $u$ changes. That is, the instanton correction ruins the
conformal invariance of Yang-Mills system at the boundary.
Far from the D-instanton position,
the metric is just the AdS background and the dilaton
is small therefore the system can be described by the type IIB supergravity.
 Notice, however, the dilaton diverge near the D-instanton position.
 There, one might expect that this system could be described by the S-dual
 of the type IIB supergravity.
In  the absence of the axion, the inversion of the coupling
($g_{s}\rightarrow 1/g_{s}$)
can be obtained by a SL(2,{\bf R}) rotation $\Omega_2$ in (\ref{slro}).
 In our case, the axion and the dilaton is tightly related by
\be
\chi +\alpha=e^{-\phi}. \label{ansatz}
\ee
Due to this relation, the axion and the dilaton transform
under a general  SL(2,{\bf R})  rotation
$\pmatrix{ a & b \cr c & d}$ into
\ba
\chi \to \chi^{\prime} &=& \frac{(ad+bc-2ac\alpha)e^{-\phi}
+(b-a\alpha)(d-c\alpha)}{(d-c\alpha)^2 + 2c(d-c\alpha)e^{-\phi}}
, \no
e^{\phi} \to e^{\phi^{\prime}} &=& (d-c\alpha)^2 e^{\phi} + 2c(d-c\alpha) .
\ea
From these, the followings can be shown easily:
\begin{itemize}
       \item  It is impossible to find a SL(2,{\bf R}) rotation which
transforms away $\chi$.

        \item  Starting from our solution,
it is impossible to get strong and weak exchange $e^\phi \to e^{-\phi}$.

        \item   With the choice $d=c\alpha$, the dilaton vanishes.
Therefore  the dyonic string, whose electric and  magnetic charge is
$ (n_e,n_m)=(a,c)$, becomes free near the D-instanton.
However, the axion blows up in this case.
\end{itemize}
All these unusual properties come from the ansatz (\ref{ansatz}).
 Consequently,  the physical
interpretation of the divergence of the  dilaton and the flatness of
the geometry near D-instanton is not very clear, yet.
Nevertheless, it is well established \cite{banks,ho,bianchi,witten5}
that the D-instanton  holographically projected
corresponds to the Yang-Mills instanton.

\vskip.5cm

What happen if we include the multi-D-instanton solutions? Multi-instanton
solution in $AdS_5 \times S^5$ can be given by
\begin{equation}
H_{-1} = h+ \sum_i \frac{ N_{-1} z_0^4 z^4}
{ [( \vec{x}-\vec{x}_i)^2 + (\vec{z}-\vec{z}_i)^2 ]^4 }
\end{equation}
For the dilute instanton gas, that is, when instantons are well separated,
multiple wormholes are generated. See figure 2. However, contrary to the
one instanton case, one can not make the string frame metric globally flat
by SL(2,{\bf R}) rotation.

\begin{figure}
\begin{center}
\includegraphics[angle=0, width=100mm]{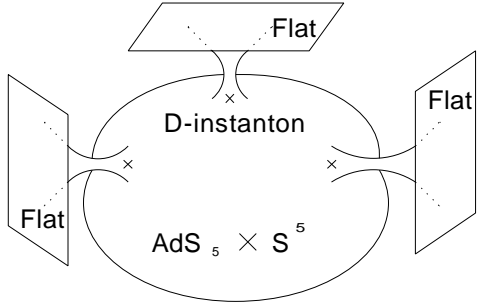}
\end{center}
\caption{ Multi-D-instanton solution generate multi-wormholes. }
\end{figure}

One may ask whether the (local) flatness near the D-instanton is general
feature of the near horizon geometry of any D-p-brane. In order to answer
this, we describe D-instanton in $AdS_3\times S^3$, which is a near horizon
geometry of the D1-D5 system. Its metric is given by
\begin{equation}
{ds_6}^2 = \frac{1}{z^2} ( {dx_0}^2 + {dx_1}^2 + d {z}^2 ) + {d\Omega_3}^2 .
\end{equation}
Here $\vec{z}$ is the four dimensional vector transverse to the D1- and
D5-brane. Now, we add D-instanton to this background.
The metric of the resulting system is given by
\begin{equation}
ds^2 = {H_{-1}}^{1/2} \left[ \frac{1}{z^2} ( {dx_0}^2 + {dx_1}^2 + d\vec{z}^2 )
\right]  \label{Din6}
\end{equation}
where $H_{-1}(\vec{x},\vec{z})$ is a harmonic function. H can be found in
similar way as before
\begin{equation}
H_{-1} = h+ \frac{ N_{-1} z_0^2 z^2}
{ [( \vec{x}-\vec{x}_0)^2 + (\vec{z}-\vec{z}_0)^2 ]^2}.
\end{equation}
Notice that the metric with the above $H_{-1}$ given in (\ref{Din6}) is not
flat near the D-instanton position. Therefore, the (local) flatness of the
geometry after the instanton correction is very special for the $AdS_5\times
S^5$ geometry.

So far, we have shown that the effect of adding D-instanton to the $AdS_5
\times S^5$ background is to make the string frame geometry locally flat
near the D-instanton position. By a $SL(2,{\bf R})$ rotation of type IIB
theory, we could  make the {\it global} geometry flat in string frame.
A subtlety on the D-instanton at the boundary   or at the horizon was
discussed.
We also found that, due to the tight relation  between the dilaton and
the axion, there is no $SL(2,{\bf R})$ element that rotates strong string
coupling  to weak  one  without making the axion ill defined.
We also  discussed the case of  $AdS_3$ as well as the instanton gases.

\vskip1cm
\noindent{\bf Acknowledgement}

\noindent We'd like to thank J. Cho, S. Hyun,
Y. Kiem and S. Nam for useful discussions and interesting questions and
thank KIAS and APCTP, where some part of the work was done,
for the hospitality and supports while SJS was visiting.
This work has been supported by the
research grant KOSEF 971-0201-001-2 and BSRI-98-2441.

\end{document}